\documentclass[cmfonts]{kapproc} % Computer Modern font calls

\usepackage{procps}

\usepackage[dvips]{graphicx}

\upperandlowercase

\kluwerbib % will produce this kind of bibliography entry:

\begin{document}

\articletitle[Surveys of extragalactic sources with {\sc
Planck}]{Surveys of extragalactic sources with {Planck}}

%\articlesubtitle{This is an Article Subtitle}

\author{G. De Zotti}
\affil{INAF - Osservatorio Astronomico di Padova, Vicolo
dell'Osservatorio 5, I-35122 Padova, Italy \\
SISSA/ISAS, Via Beirut 2-4, I-34014 Trieste, Italy}
\author{C. Burigana}
\affil{CNR-INAF/IASF, Sezione di Bologna, Via Gobetti 101, I-40129
Bologna, Italy}
\author{M. Negrello, S. Tinti and R. Ricci}
\affil{SISSA/ISAS, Via Beirut 2-4, I-34014 Trieste, Italy}
\author{L. Silva}
\affil{INAF--Osservatorio Astronomico di Trieste, Via G.B. Tiepolo
11, I-34131 Trieste, Italy}
\author{J. Gonzalez-Nuevo and L. Toffolatti}
\altaffiltext{5}{Departamento de Fisica, Universidad de Oviedo, c.
Calvo Sotelo s/n, 33007 Oviedo, Spain}

% optional abstract
\begin{abstract}
 Although the primary goal of ESA's {\sc Planck} mission is to produce high
 resolution maps of the temperature and polarization anisotropies of the Cosmic
 Microwave Background (CMB), its high-sensitivity all-sky surveys of extragalactic sources
 at 9 frequencies in the range 30--860 GHz will constitute a major
 aspect of its science products. In particular, {\sc Planck}
 surveys will provide key information on several highly
 interesting radio source populations, such as Flat Spectrum Radio
 Quasars (FSRQs), BL Lac objects, and, especially, extreme GHz
 Peaked Spectrum (GPS) sources, thought to correspond to the very
 earliest phases of the evolution of radio sources. Above 100 GHz,
 {\sc Planck} will provide the {f}{i}rst all-sky surveys, that are expected
 to supply rich samples of highly gravitationally ampli{f}{i}ed dusty
 proto-galaxies and large samples of candidate proto-clusters at
 $z\simeq 2$--3, thus shedding light on the evolution of large
 scale structure across the cosmic epoch when dark energy should
 start dominating the cosmic dynamics.

\end{abstract}

% optional keywords

\def\lsim{\,\lower2truept\hbox{${<\atop\hbox{\raise4truept\hbox{$\sim$}}}$}\,}
\def\gsim{\,\lower2truept\hbox{${> \atop\hbox{\raise4truept\hbox{$\sim$}}}$}\,}

\section{Introduction}

The {\sc Planck} satellite will carry our high sensitivity all sky
surveys at 9 frequencies in the poorly explored range 30--860 GHz
(see the contribution by J. Tauber, this volume). At low
frequencies, {\sc Planck} will go several times deeper (and will
detect about ten times more sources) than WMAP, that has provided
the {f}{i}rst all-sky surveys at frequencies of tens of GHz,
comprising about 200 objects (Bennett et al. 2003).

Above 100 GHz, {\sc Planck} surveys will be the {f}{i}rst and will
remain the only all sky surveys available for many years to come.
They will {f}{i}ll an order of magnitude gap in our knowledge of
the spectrum of bright extragalactic sources and may discover new
populations, not represented, or not recognized, in lower or
higher frequencies surveys.

Rather than presenting a comprehensive review of the expected
scienti{f}{i}c results from {\sc Planck} measurements of
extragalactic sources (see, e.g., De Zotti et al. 1999) we will
focus on a couple of frequencies, one of the Low Frequency
Instrument, namely 30 GHz, and one of the High Frequency
Instrument, namely 350 GHz. The relatively shallow but all-sky
{\sc Planck} surveys will be ideal to study populations which are
both very powerful at mm/sub-mm wavelengths, and very rare, such
as radio sources with inverted spectra up to $\ge 30\,$GHz
[extreme GHz Peaked Spectrum (GPS) sources or High Frequency
Peakers (HFP)], thought to be the most recently formed and among
the most luminous radio sources, and ultra-luminous dusty
proto-spheroidal galaxies, undergoing their main and huge episode
of star formation at typical redshifts $\ge 2$ (Granato et al.
2001, 2004). And {\sc Planck} will observe such sources with an
unprecedented frequency coverage.

To estimate the detection limit, and the number of detectable
sources, we need to take into account, in addition to the
instrument noise, the {f\-l}uctuations due to Galactic emissions,
to the Cosmic Microwave Background (CMB), and to extragalactic
sources themselves. These {f\-l}uctuations will be brie{f}{l}y
reviewed in Section 2, while in Sections 3 and 4 we will discuss
the expected impact of {\sc Planck} data on our understanding of
HFPs and of ultra-luminous proto-spheroidal galaxies,
respectively. Our main conclusions are summarized in Sect. 5.

\begin{figure*}[t]
%\vskip-5cm
\centerline{
\protect\includegraphics[width=12cm,height=10cm]{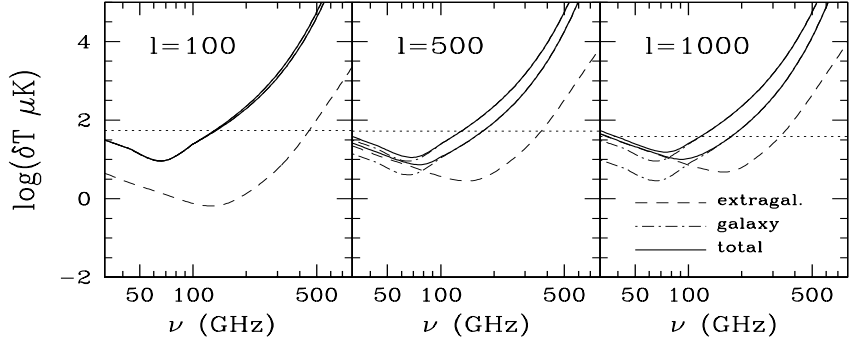}}
\vskip-5cm \caption{Galactic (dot-dashed) and extragalactic
(dashed) contributions to the power spectrum of foreground
{f\-l}uctuations, compared with the CMB (dotted horizontal line)
for three values of the multipole number $\ell$. The solid lines
show the sum, in quadrature, of the two contributions. At
$\ell=100$ the dot-dashed line essentially coincides with the
solid line; the two lines largely overlap at high frequencies also
for higher $\ell$'s. The Galactic contributions are averages for
$|b|\ge 20^\circ$, after having applied the Kp0 mask which include
the point source removal, and comprise synchrotron, free-free and
thermal dust emissions, whose power spectra are normalized to the
K-band (22.8 GHz), V-band (60.8 GHz), and W-band (93.5 GHz) WMAP
data, respectively (where each component is best measured). The
extrapolation in frequency has been done adopting, for free-free,
the antenna temperature spectral index ($T_A\propto \nu^\beta$)
$\beta_{\rm ff}=-2.15$, and for synchrotron the expression
proposed by Jackson \& Wall (2002) for low-luminosity radio
sources ($\log S_\nu = \hbox{const} - 0.6424\log(\nu) -
0.0692(\log(\nu)^2)$, with $\nu$ in GHz); this formula, which
allows for the high-frequency steepening of the synchrotron
spectrum due to electron energy losses, is consistent with the
steepening observed in WMAP data (Fig. 9 of Bennett et al. 2003).
As for thermal dust we have considered two cases: $\beta_d=2.2$,
the best {f}{i}t value of Bennett et al. (2003), and the more
usual value $\beta_d=2$. With these spectra, an additional
component (spinning dust?) is necessary to account for the
foreground signal detected by WMAP particularly in the Q-band
(40.7 GHz); the solid lines include this component. Power spectra
at $\ell =100$ were derived directly from WMAP data. At higher
$\ell$'s we assume $C_\ell = C_{100} (\ell/100)^{-\gamma}$ with
$\gamma = 2$ or $\gamma = 3$. The upper dot-dashed curve
corresponds to $\gamma = 2$ and $\beta_d=2.2$, the lower one to
$\gamma = 3$ and $\beta_d=2$. The dashed curve includes, summed in
quadrature, the contributions of all classes of extragalactic
sources, based on models by De Zotti et al. (2004), including
canonical radio sources, starburst galaxies, proto-spheroidal
galaxies and Sunyaev-Zeldovich effects. The effect of clustering
of proto-spheroidal galaxies has been taken into account as in
Negrello et al. (2004a). }\label{foreground}
\end{figure*}

\section{Power spectra of foreground emissions}

For a very high sensitivity experiment, like {\sc Planck}, the
main limitation to the capability of mapping the CMB is set by
contamination by astrophysical sources (``foregrounds''), while
CMB {f\-l}uctuations may be the highest ``noise'' source for the
study of astrophysical emissions at mm wavelengths. The most
intense foreground source is our own Galaxy. Because of the
different power spectra of the various emission components, the
frequency of minimum foreground {f\-l}uctuations depends to some
extent on the angular scale (see Fig.~\ref{foreground}, where
$\delta T$ are {f\-l}uctuations of the CMB thermodynamic
temperature, in $\mu$K, related to the power spectrum $C_\ell$ by
$\delta T = [\ell(\ell+1)C_\ell/(2\pi)]^{0.5}(e^x-1)^2/(x^2 e^x)$,
with $x=h\nu/kT_{\rm CMB}$). So long as diffuse Galactic emissions
dominate the {f\-l}uctuations ($\theta \gsim 30'$; see De Zotti et
al. 1999), they have a minimum in the 60--80 GHz range (depending
also on Galactic latitude; cf. Bennett et al. 2003).

But the power spectra of diffuse Galactic emissions decline rather
steeply with increasing multipole number (or decreasing angular
scale). Thus, on small scales, {f\-l}uctuations due to
extragalactic sources, whose Poisson contribution has a
white-noise power spectrum (on top of which we may have a,
sometimes large, clustering contribution) take over, even though
their integrated emission is below the Galactic one. At high
frequencies, however, Galactic dust may dominate {f\-l}uctuations
up to $\ell$ values of several thousands. Unlike the relatively
quiescent Milky Way, the relevant classes of extragalactic sources
have strong nuclear radio activity or very intense star formation,
or both. Thus, although in many cases their SEDs are qualitatively
similar to that of the Milky Way, there are important quantitative
differences. In particular, dust in active star forming galaxies
is signi{f\-i}cantly hotter and the radio to far-IR intensity
ratio of the extragalactic background is much higher than that of
the Milky Way. Both factors, but primarily the effect of radio
sources, cooperate to move the minimum of the SED to 100--150 GHz.

\begin{figure}[t]
\centerline{\includegraphics[width=5.5cm,height=5cm]{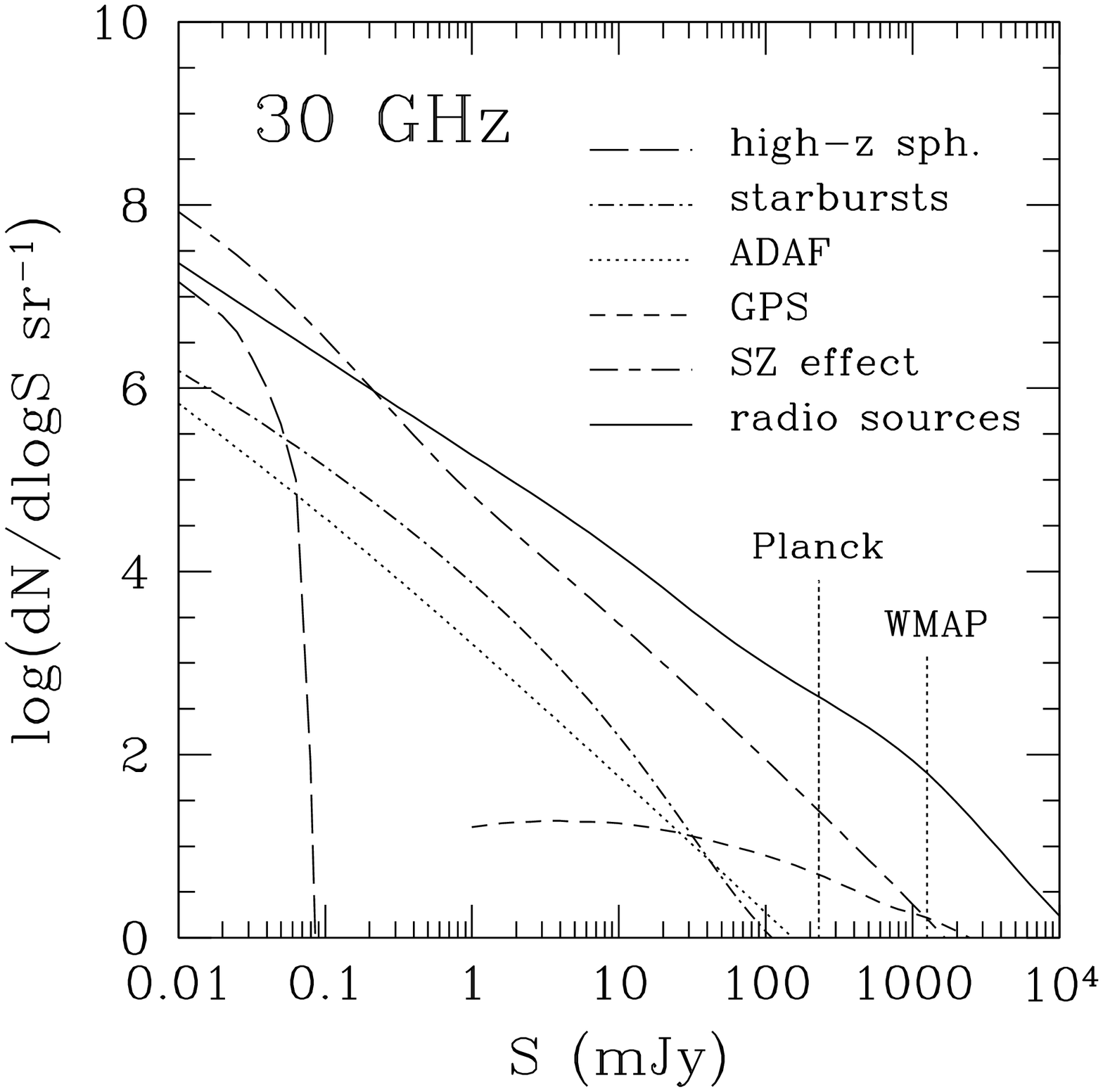}
\includegraphics[width=5.5cm,height=5cm]{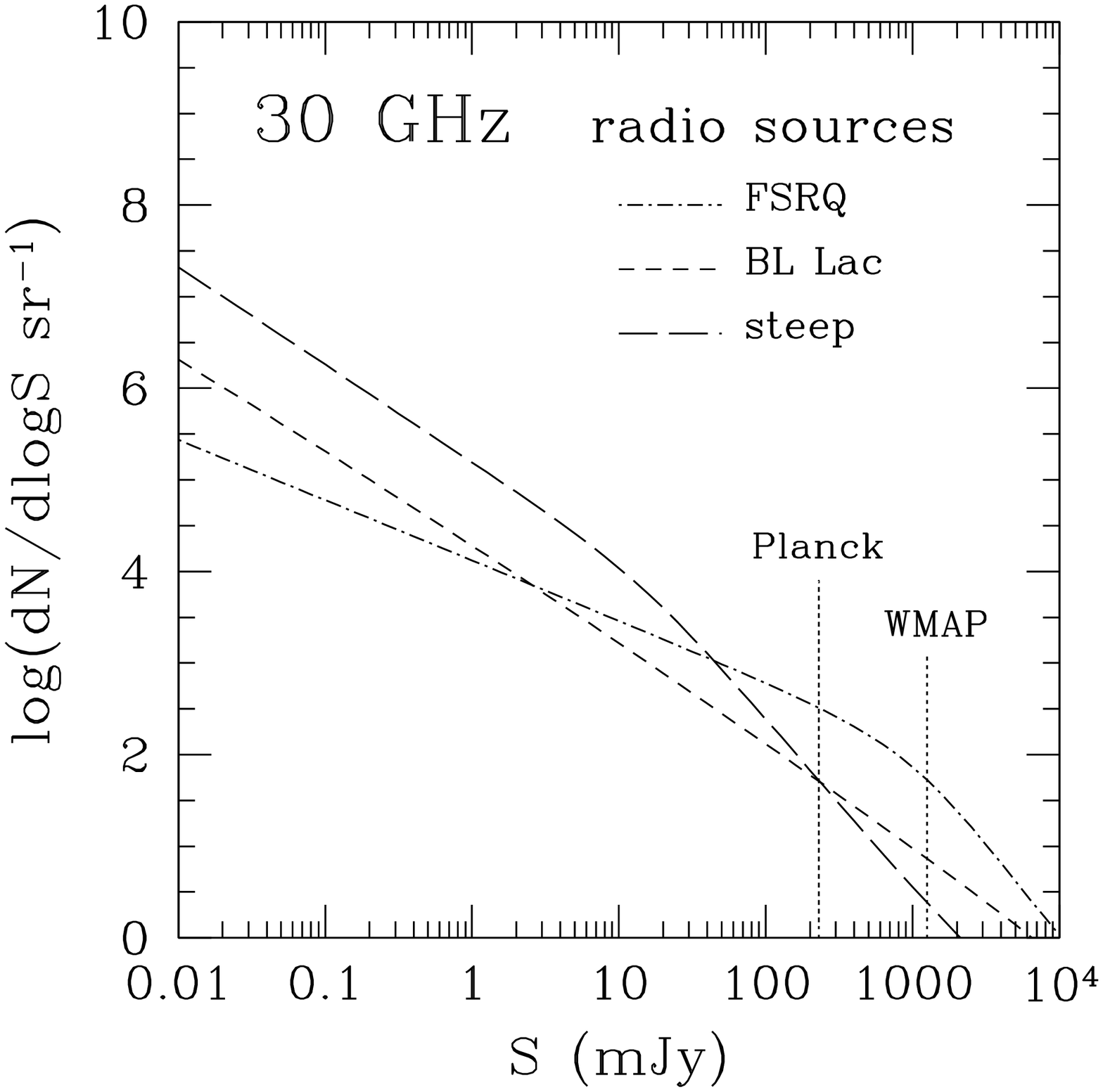}}
\caption{Predicted 30 GHz differential counts. The left-hand panel
shows the counts of all the main populations (see De Zotti et al.
2004 for details). The right-hand panel details the contributions
of three sub-classes of canonical radio sources: FSRQs, BL Lac
objects, and steep-spectrum sources } \label{30GHzcounts}
\end{figure}

\begin{figure*}
\centerline{\includegraphics[width=5.5cm,height=5cm]{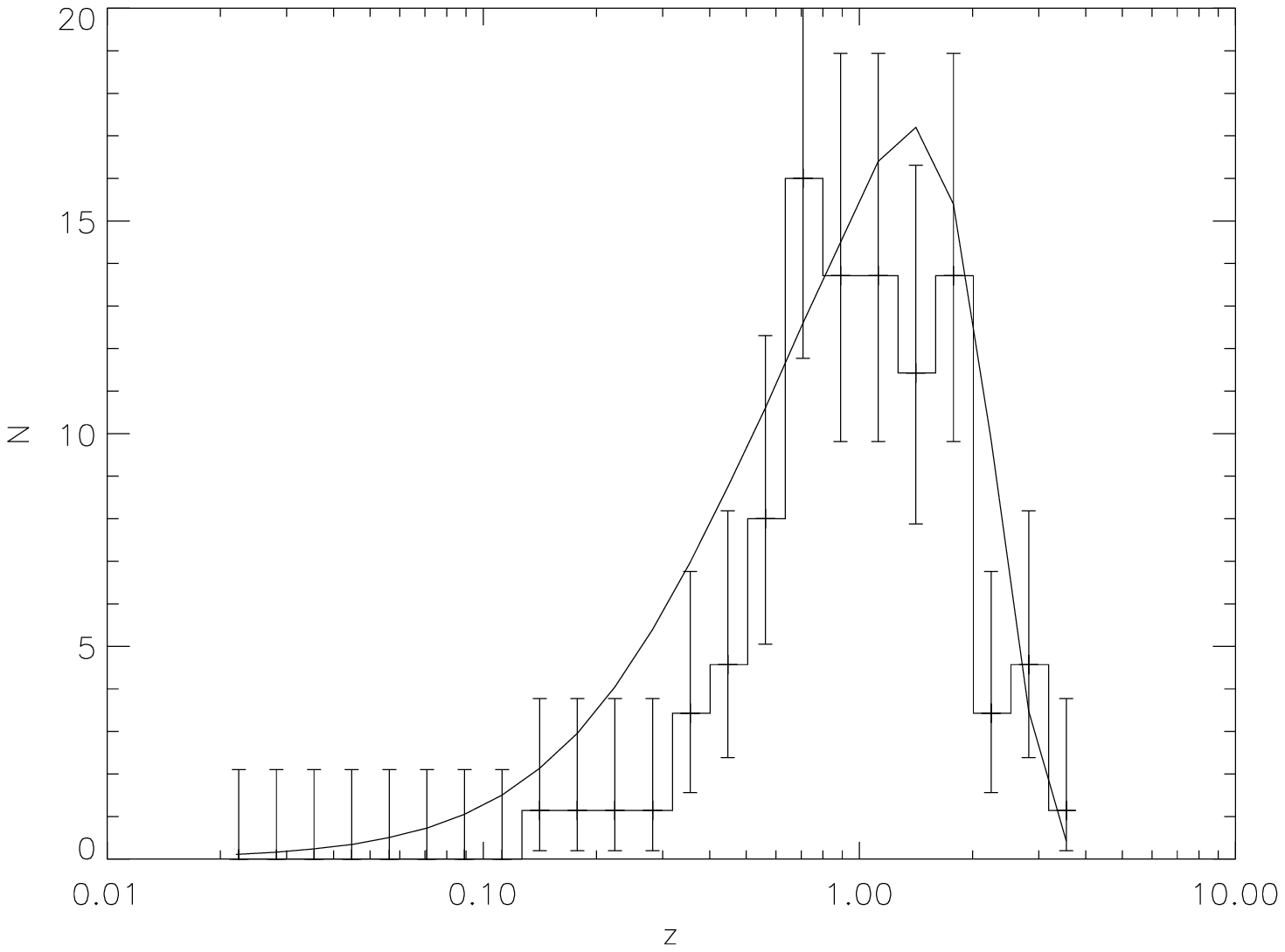}
\includegraphics[width=5.5cm,height=5cm]{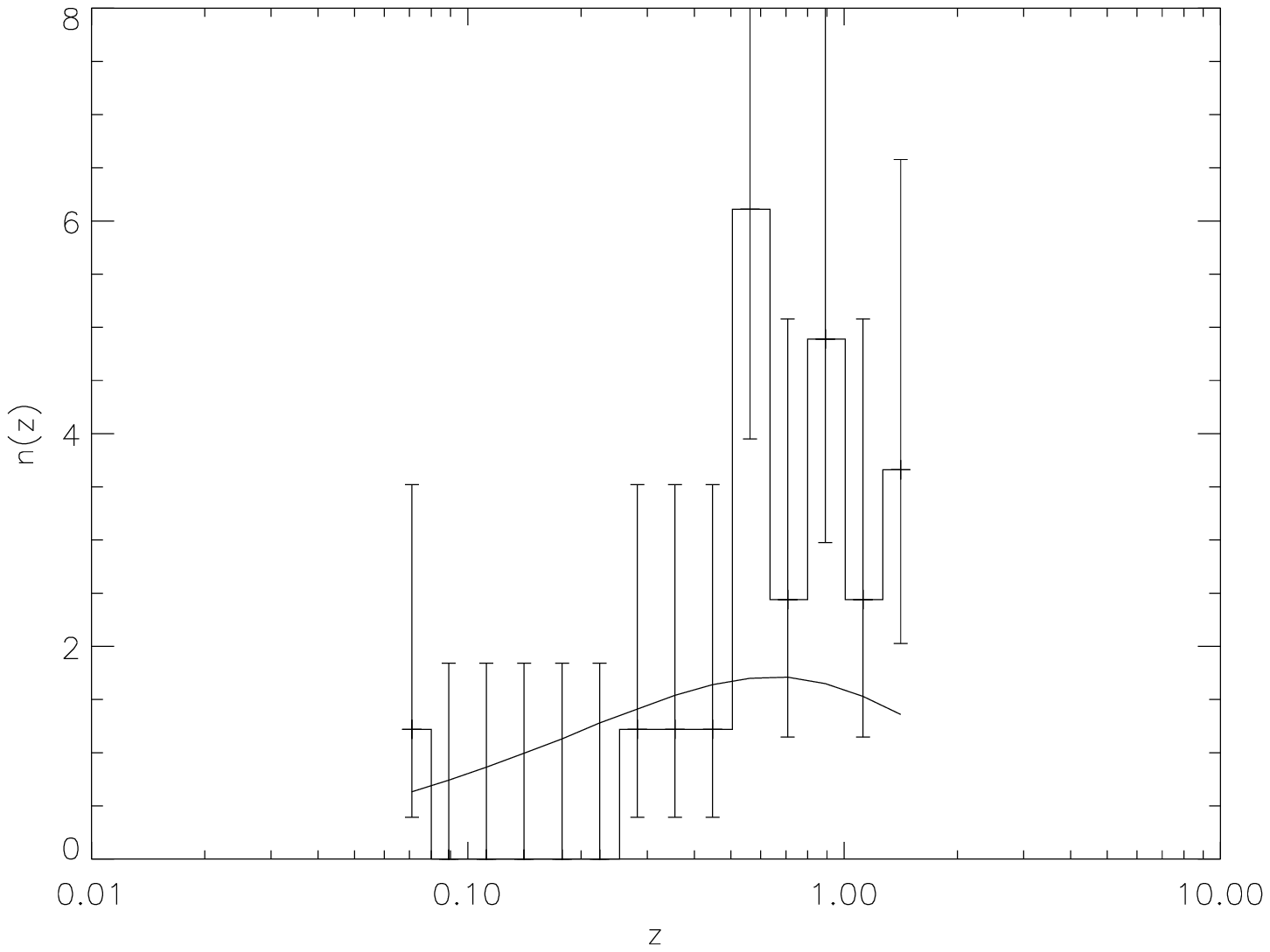}}
\caption{Redshift distributions of WMAP FSRQs (left-hand panel)
and BL Lacs (right-hand panel) compared with the model by De Zotti
et al. (2004, solid line). }\label{WMAP_FSRQ_BLLacs}
\end{figure*}

\section{30 GHz counts}

Figure~\ref{30GHzcounts} provides a synoptic view of the
contributions of different source classes to the global counts of
extragalactic sources. Shallow surveys, such as those by WMAP and
{\sc Planck}, mostly detect canonical radio sources.  As shown by
the right-hand panel of Fig.~\ref{30GHzcounts}, detected sources
will be mostly {f}{l}at-spectrum radio quasars (FSRQs), while the
second more numerous population are BL Lac objects. {\sc Planck}
will detect about ten times more sources than WMAP, thus allowing
a substantial leap forward in the understanding of evolutionary
properties of both populations at high frequencies, only weakly
constrained by WMAP data (Fig.~\ref{WMAP_FSRQ_BLLacs}).

{\sc Planck} will also provide substantial complete samples of
sources not (yet) represented in the WMAP catalog, such as
Sunyaev-Zeldovich (1972) signals and extreme GPS sources or HFPs
(Dallacasa et al. 2000).

GPS sources are powerful ($\log P_{\rm 1.4\, GHz} \gsim
25\,\hbox{W}\,\hbox{Hz}^{-1}$), compact ($\lsim 1\,$kpc) radio
sources with a convex spectrum peaking at GHz frequencies. It is
now widely agreed that they correspond to the early stages of the
evolution of powerful radio sources, when the radio emitting
region grows and expands within the interstellar medium of the
host galaxy, before plunging in the intergalactic medium and
becoming an extended radio source (Fanti et al. 1995; Readhead et
al.  1996; Begelman 1996; Snellen et al. 2000). Conclusive
evidence that these sources are young came from measurements of
propagation velocities. Velocities of up to $\simeq 0.4c$ were
measured, implying dynamical ages $\sim 10^3$ years (Polatidis et
al. 1999; Taylor et al. 2000; Tschager et al. 2000). The
identi{f}{i}cation and investigation of these sources is therefore
a key element in the study of the early evolution of radio-loud
AGNs.

There is a clear anti-correlation between the peak (turnover)
frequency and the projected linear size of GPS sources. Although
this anti-correlation does not necessarily de{f}{i}ne the
evolutionary track, a decrease of the peak frequency as the
emitting blob expands is indicated. Thus high-frequency surveys
may be able to detect these sources very close to the moment when
they turn on. The self-similar evolution models by Fanti et al.
(1995) and Begelman (1996) imply that the radio power drops as the
source expands, so that GPS's evolve into lower luminosity radio
sources, while their luminosities are expected to be very high
during the earliest evolutionary phases, when they peak at high
frequencies. De Zotti et al. (2000) showed that, with a suitable
choice of the parameters, this kind of models may account for the
observed counts, redshift and peak frequency distributions of the
samples then available. The models by De Zotti et al. (2000)
imply, for a maximum rest-frame peak frequency $\nu_{p,i}
=200\,$GHz, about 10 GPS quasars with $S_{30{\rm GHz}} > 2\,$Jy
peaking at $\geq 30\,$GHz over the 10.4 sr at $|b| >10^\circ$.

Although the number of {\it candidate} GPS quasars (based on the
spectral shape) in the WMAP survey is consistent with such
expectation, when data at additional frequencies (Trushkin 2003)
are taken into account the GPS candidates look more blazars caught
during a {f}{l}are optically thick up to high frequencies.
Furthermore, Tinti et al. (2004) have shown that most, perhaps two
thirds, of the quasars in the sample of HFP candidates selected by
Dallacasa et al. (2000) are likely blazars.

Thus, WMAP data are already providing strong constraints on the
evolution of HFPs. {\sc Planck} will substantially tighten such
constraints and may allow us to directly probe the earliest phases
(ages $\sim 100\,$yr) of the radio galaxy evolution, hopefully
providing hints on the still mysterious mechanisms that trigger
the radio activity.

We note, in passing, that contrary to some claims, we do not
expect that {\sc Planck} can detect the late phases of the AGN
evolution, characterized by low accretion/radiative ef{f}{i}ciency
(ADAF/ADIOS sources).

At faint {f\-l}ux densities, other populations come out and are
expected to dominate the counts. In addition to SZ effects, we
have active star-forming galaxies, seen either through their radio
emission, or through their dust emission, if they are at
substantial redshift. The latter is the case for the sub-mm
sources detected by the SCUBA surveys if they are indeed at high
redshifts (see below).

Such sources may be relevant in connection with the interpretation
of the excess signal on arc-minute scales detected by CBI (Mason
et al. 2003; Readhead et al. 2004) and BIMA (Dawson et al. 2002)
experiments at 30 GHz, particularly if, as discussed below, they
are highly clustered, so that their contribution to
{f\-l}uctuations is strongly super-Poissonian (Toffolatti et al.
2004). In fact, to abate the point source contamination of the
measured signals, the CBI and BIMA groups could only resort to
existing or new low frequency surveys. But the dust emission is
undetectable at low frequencies. Although our reference model
(Granato et al. 2004), with its relatively warm dust temperatures
yielded by the code GRASIL (Silva et al. 1998), imply dusty galaxy
contributions to small scale {f\-l}uctuations well below the
reported signals, the (rest-frame) mm emission of such galaxies is
essentially unknown and may be higher than predicted, e.g. in the
presence of the extended distribution of cold dust advocated by
Kaviani et al. (2003) or of a widespread mm excess such as that
detected in several Galactic clouds (Dupac et al. 2003) and in
NGC1569 (Galliano et al. 2003). This is another instance of the
importance of a multifrequency approach, like {\sc Planck}'s,
capable of keeping under control all the relevant emission
components, with their different emission spectra.

\begin{figure}
\centerline{\includegraphics[width=11.5cm,height=10.5cm]{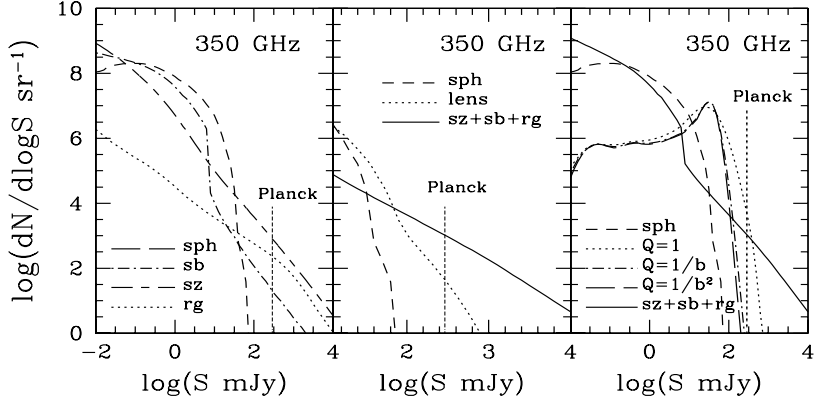}
} \vskip-5cm \caption{Left-hand panel: contributions of different
populations to the 350 GHz counts. Central panel: effect of
lensing on counts of proto-spheroidal galaxies. Right-hand panel:
estimated counts of ``clumps'' of proto-spheroids observed with
{\sc Planck} resolution. } \label{lensplusclust}
\end{figure}

\section{350 GHz counts}

The 350 GHz counts of extragalactic sources have been determined
in the range from $\simeq 10\,$mJy to $\simeq 0.25\,$mJy by
surveys with the SCUBA camera, covering small areas of the sky
(overall, a few tenths of a square degree). These surveys have led
to the discovery of very luminous high-$z$ galaxies, with
star-formation rates $\sim 10^3\,M_\odot$/yr, a result
con{f}{i}rmed by 1.2mm surveys with MAMBO on the IRAM 30m
telescope. These data proved to be extremely challenging for
semi-analytic galaxy formation models, and have indeed forced to
reconsider the evolution of baryons in dark matter halos.

The bright portion of observed counts appears to be declining
steeply with increasing {f\-l}ux density, probably re{f\-l}ecting
the exponential decline of the dark-halo mass function at large
masses implied by the Press \& Schechter formula, so that one
would conclude that {\sc Planck} cannot do much about these
objects, but rather detect brighter sources such as blazars and
relatively local star-forming galaxies, or SZ signals. There are,
however, two important effects to be taken into account, that may
change this conclusion: gravitational lensing and clustering.

We refer here to the model by Granato et al. (2004) according to
which SCUBA sources are large spheroidal galaxies in the process
of forming most of their stars. Forming spheroidal galaxies, being
located at relatively high redshift, have a substantial optical
depth for gravitational lensing, and the effect of lensing on
their counts is strongly ampli{f\-i}ed by the steepness of the
counts. This is illustrated by the left-hand panel of
Fig.~\ref{lensplusclust}, based on calculations by Perrotta et al.
(2003). Strong lensing is thus expected to bring a signi{f\-i}cant
number of high-$z$ forming spheroids above the estimated {\sc
Planck} $5\sigma$ detection limit.

If indeed SCUBA galaxies are massive spheroidal galaxies at high
$z$, they must be highly biased tracers of the matter
distribution, and must therefore be highly clustered. There are in
fact several, although tentative, observational indications of
strong clustering with comoving radius $r_0 \simeq
8\hbox{h}^{-1}\,$Mpc (Smail et al. 2004; Blain et al. 2004;
Peacock et al. 2000), consistent with theoretical expectations.

But if massive spheroidal proto-galaxies live in strongly
over-dense regions, low resolution experiments like {\sc Planck}
unavoidably measure not the {f\-l}ux of individual objects but the
sum of {f\-l}uxes of all physically related sources in a
resolution element.

This is an aspect of the ``source confusion" problem, whereby the
observed {f\-l}uxes are affected by unresolved sources in each
beam. The problem was extensively investigated in the case of a
Poisson distribution, particularly by radio astronomers (Scheuer
1957, Murdoch et al. 1973, Condon 1974, Hogg \& Turner 1998). The
general conclusion is that unbiased {f\-l}ux measurements require
a $S/N \ge 5$.

Not much has been done yet on confusion in the presence of
clustering (see however Hughes \& Gaztanaga 2000). The key
difference is that, for a Poisson distribution, a bright source is
observed on top of a background of unresolved sources that may be
either above or below the all-sky average, while in the case of
clustering, sources are preferentially found in over-dense
regions.

Clearly, the excess signal (over the {f\-l}ux of the brightest
source in the beam) depends on the angular resolution. For a
standard $\xi(r) = (r/r_0)^{-1.8}$ the mean clustering
contribution is $\propto r_0^{1.8} r_{\rm beam}^{1.2}$. The {\sc
Planck} beam at this frequency corresponds to a substantial
portion of the typical clustering radius at $z\simeq 2$--3, so
that {\sc Planck} will actually measure a signi{f\-i}cant fraction
of the {f\-l}ux of the clump, which may be substantially larger
than the {f\-l}ux of any member source. The effect on counts
depends on the joint distribution of over-densities and of $M/L$
ratios. The former depends on both the two- and the three-point
correlation function, while the latter depends on the luminosity
function.

Preliminary estimates of the distribution of excess luminosities
due to clustering around bright sources have been obtained by
Negrello et al. (2004b). The right-hand panel of
Fig.~\ref{lensplusclust} shows the estimated counts of clumps
observed with {\sc Planck} resolution for three models for  the
evolution of the coef{f}{i}cient $Q$ of the three-point
correlation function. Obviously {\sc Planck} can provide
information only on the brightest clumps, and, except in the
extreme case of $Q=1$ at all cosmic times, the clumps will only
show up as $< 5\sigma$ {f\-l}uctuations. On the other hand, such
{f\-l}uctuations will provide a rich catalogue of candidate
proto-clusters at substantial redshifts (typically at $z\simeq
2$--3), very important to investigate the formation of large scale
structure and, particularly, to constrain the evolution of the
dark energy thought to control the dynamics of the present day
universe.

\section{Conclusions}

Although extragalactic surveys are not the primary goal of the
mission, {\sc Planck}  will provide unique data for several
particularly interesting classes of sources. Examples are the
FSRQs, BL Lac objects, but especially extreme GPS sources that may
correspond to the earliest phases of the life of radio sources,
and proto-spheroidal galaxies. Thus {\sc Planck} will investigate
not only the origin of the universe but also the origin of radio
activity and of galaxies. Sub-mm surveys will provide large
samples of candidate proto-clusters, at $z\simeq 2$--3, shedding
light on the evolution of the large scale structure (and in
particular providing information on the elusive three-point
correlation function) and of the dark energy, across the cosmic
epoch when it is expected to start dominating the cosmic dynamics.

\begin{acknowledgments}
Work supported in part by MIUR through a PRIN grant and by ASI.
\end{acknowledgments}

% The endnotes section will be placed here.

%\theendnotes

%\end{article}

\begin{chapthebibliography}{1}

\bibitem{}
Begelman, M.C. (1996). In ``Cygnus A -- Study of a Radio Galaxy'',
C.L. Carilli \& D.E. Harris eds., Cambridge University Press, p.
209

\bibitem{}
Bennett, C.L., et al. (2003).  ApJS, 148, 97.

\bibitem{}
Blain, A. W., S.C. Chapman, I. Smail, and R. Ivison (2004). ApJ,
611, 725

\bibitem{}
Condon, J.J. (1974). ApJ, 188, 279.

\bibitem{}
Dallacasa, D., C. Stanghellini, M. Centonza, and R. Fanti (2000).
A\&A, 363, 887.

\bibitem{}
Dawson, K.S., et al. (2002). ApJ, 581, 86.

\bibitem{}
De Zotti, G., et al. (1999). In L. Maiani, F. Melchiorri and N.
Vittorio, editors, {\em Proceedings of the EC-TMR Conference ``3K
Cosmology'', AIP CP}, 476, 204.

\bibitem{}
De Zotti, G., G.L. Granato, L. Silva, D. Maino, and L. Danese
(2000). A\&A, 354, 467.

\bibitem{}
De Zotti, G., R. Ricci, D. Mesa, L. Silva, P. Mazzotta, L.
Toffolatti, and J. Gonz\'alez-Nuevo (2004). A\&A, in press

\bibitem{}
Dupac, X., et al. (2003). A\&A, 404, L11.


\bibitem{}
Fanti, C., et al. (1995). A\&A, 302, 317.

\bibitem{}
Galliano, F., et al. (2003). A\&A, 407, 159.

\bibitem{}
Granato, G.L., L. Silva, P. Monaco, P. Panuzzo, P. Salucci, G. De
Zotti, and L. Danese (2001).  MNRAS, 324, 757.

\bibitem{}
Granato, G.L., G. De Zotti, L. Silva, A. Bressan, and L. Danese
(2004). ApJ, 600, 580.

\bibitem{}
Hogg, D.W., and E.L. Turner (1998). PASP, 110, 727.

\bibitem{}
Hughes, D.H., and E. Gaztanaga (2000). In {\em Star formation from
the small to the large scale}, proc. 33rd ESLAB symp., F. Favata,
A. Kaas, and A. Wilson eds., ESA SP 445, p. 29.

\bibitem{}
Jackson, C.A., and J.V. Wall (2001). In {\em Particles and Fields
in Radio Galaxies}, ASP Conf. Proc. 250, R.A. Laing and K.M.
Blundell eds., p. 400.

\bibitem{}
Kaviani, A., M.G. Haehnelt, and G. Kauffmann (2003). MNRAS, 340,
739.

\bibitem{}
Mason, B.S., et al. (2003). ApJ, 591, 540.

\bibitem{}
Murdoch, H.S., D.F. Crawford, and D.L. Jauncey (1973). ApJ, 183,
1.

\bibitem{} Negrello, M., M. Magliocchetti, L. Moscardini, G. De Zotti,
G.L. Granato, and L. Silva (2004a). MNRAS, 352, 493.

\bibitem{}
Negrello, M., M. Magliocchetti, L. Moscardini, G. De Zotti, and L.
Danese (2004b). MNRAS, submitted

\bibitem{}
Peacock, J.A., et al. (2000). MNRAS, 318, 535.

\bibitem{}
Perrotta, F., M. Magliocchetti, C. Baccigalupi, M. Bartelmann, G.
De Zotti, G.L. Granato, L. Silva, and L. Danese (2003). MNRAS,
338, 623.

\bibitem{}
Polatidis, A., et al. (1999). NewAR, 43, 657.

\bibitem{}
%Readhead, A.C.S., et al. (1996). ApJ, 460, 634.
Readhead, A.C.S., G.B. Taylor, T.J. Pearson, and P.N. Wilkinson
(1996). ApJ, 460, 634.

\bibitem{}
Readhead, A.C.S., et al. (2004). ApJ, 609, 498.

\bibitem{}
Scheuer, P.A.G. (1957). Proc. Cambridge Phil. Soc., 53, 764.

\bibitem{}
Silva, L., G.L. Granato, A. Bressan, and L. Danese (1998). ApJ,
509, 103.

\bibitem{}
Smail, I., S.C. Chapman, A.W. Blain, and R.J. Ivison (2004). In
``The Dusty and Molecular Universe: A Prelude to Herschel and
ALMA'', A. Wilson ed.,  ESA Conf. Ser., p. 16.

\bibitem{}
Snellen, I.A.G., et al. (2000). MNRAS, 319, 445.

\bibitem{}
Sunyaev, R. A., and Ya.B. Zeldovich (1972). Comm. Ap. Space Phys.,
4, 173.

\bibitem{}
Taylor, G.B., J.M. Marr, T.J. Pearson, and A.C.S. Readhead (2000).
ApJ, 541, 112.

\bibitem{}
Tinti, S., D. Dallacasa, G. De Zotti, A. Celotti, and C.
Stanghellini (2004).  A\&A, in press.

%\bibitem{}
%Toffolatti, L., F. Argueso Gomez, G. de Zotti, P. Mazzei, A.
%Franceschini, L. Danese, and C. Burigana (1998). MNRAS, 297, 117.

\bibitem{}
Toffolatti, L., J. Gonz\'alez-Nuevo, M. Negrello, G. De Zotti, L.
Silva, G.L. Granato, and F. Arg\"ueso (2004). A\&A, submitted

\bibitem{}
Trushkin, S.A. (2003). Bull. Spec. Astrophys. Obs. N. Caucasus,
55, 90, astro-ph/0307205.

\bibitem{}
Tschager, W., et al. (2000). A\&A, 360, 887.

\end{chapthebibliography}
\end{document}